\documentclass[11pt]{article}
\usepackage{epsfig}
\usepackage{mathtools}
\usepackage{amsfonts}
\usepackage{amssymb}
\usepackage{amsmath}
\usepackage{listings}

\usepackage{color}
\usepackage{slashed}
\usepackage{cite}
\usepackage{caption}
\usepackage{subcaption}

\usepackage{mmacells}

\usepackage{geometry} 
\usepackage{graphicx}
\graphicspath{{pictures/}}

\usepackage{multirow}           
\usepackage{array}              

\def\ep  {\varepsilon}
\def\E4  {\mathcal{E}_4}

\newcommand{\OO}{\mathcal{O}}

\newcommand{\bx}{{\bf x}}

\newcommand{\bm}{{\bf m}}
\newcommand{\ba}{{\bf a}}
\newcommand{\bb}{{\bf b}}
\newcommand{\bc}{{\bf c}}
\newcommand{\be}{{\bf e}}

\newcommand{\bk}{{\bf k}}

\begin{document}

\begin{center}

\vspace{3cm}

{\bf \Large High-precision numerical evaluation of Lauricella functions} \vspace{1cm}

{\large M.A. Bezuglov$^{1}$, B.A. Kniehl$^{1}$, A.I. Onishchenko$^{2,3}$ and O.L. Veretin$^{4}$}\vspace{0.5cm}

{\it $^1$ II.~Institut f\"ur Theoretische Physik, Universit\"at Hamburg, Hamburg, Germany \\	
$^2$ Bogoliubov Laboratory of Theoretical Physics, Joint
Institute for Nuclear Research, Dubna, Russia, \\
$^3$ Budker Institute of Nuclear Physics, Novosibirsk, Russia} \\
$^4$ Institut f\"ur Theoretische Physik, Universit\"at Regensburg, Regensburg, Germany
\vspace{1cm}
\end{center}

\begin{abstract}
	
We present a method for high-precision numerical evaluations of Lauricella functions, whose indices are linearly dependent on some parameter $\varepsilon$, in terms of their Laurent series expansions at zero. This method is based on finding analytic continuations of these functions in terms of Frobenius generalized power series. Being one-dimensional, these series are much more suited for high-precision numerical evaluations than multi-dimensional sums arising in approaches to analytic continuations based on re-expansions of hypergeometric series or Mellin--Barnes integral representations. To accelerate the calculation procedure further, the $\varepsilon$ dependence of the result is reconstructed from the evaluations of given Lauricella functions at specific numerical values of $\varepsilon$, which, in addition, allows for efficient parallel implementation. The method has been implemented in the \texttt{PrecisionLauricella} package, written in Wolfram Mathematica language. 
\end{abstract}

\begin{center}
\end{center}

\newpage

\tableofcontents{}\vspace{0.5cm}

\renewcommand{\theequation}{\thesection.\arabic{equation}}

\section{Introduction}
\label{sec:Introduction}

\begin{figure}[ht]
	\centering
	\includegraphics[width=0.8\textwidth]{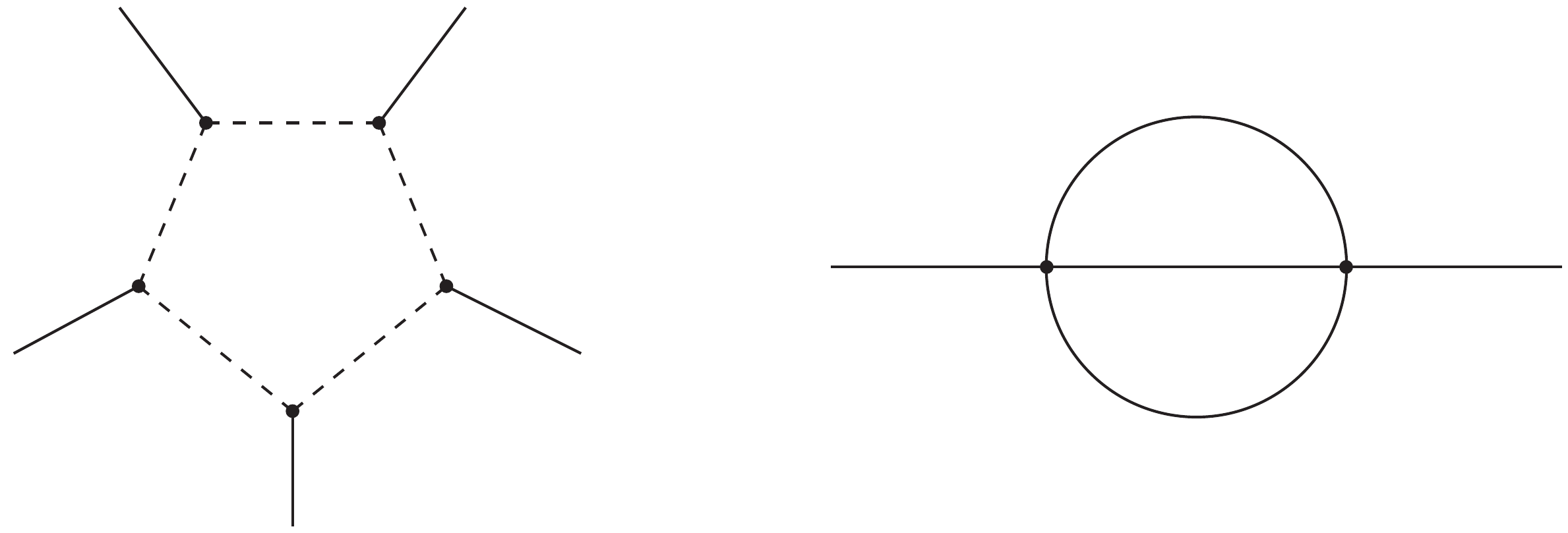}
	\caption{Massless pentagon
integral on the left and two-loop sunset on the right. Dashed lines denote massless propagators, and thick lines represent massive propagators.}
	\label{fig:exSimplePheynman}
\end{figure}

Hypergeometric functions of one or several variables frequently appear in different areas of physics and mathematics, especially in quantum field theory and computations of Feynman integrals. Various methods have been used to express Feynman integrals in terms of hypergeometric functions, such as the Mellin--Barnes method,\footnote{See Refs.~\cite{Weinzierl:2022eaz,Dubovyk:2022obc,smirnov2006feynman} for introduction and references to original works.} the DRA method \cite{Tarasov:2006nk,  Lee:2009dh, Lee:2012hp}, the method of functional equations \cite{Tarasov:2022clb}, the exact Frobenius method \cite{Bezuglov:2022npo, Bezuglov:2021tax, Blumlein:2021hbq} and, of course, the Gelfand--Kapranov--Zelevinsky (GKZ) approach to Feynman integrals\footnote{See Refs.~\cite{Matsubara-Heo:2023ylc,Vanhove:2018mto}, for an introduction and general overview.} \cite{GKZ1,GKZ2,GKZ3,GKZ4,GKZ5,beukers2013monodromy,Kalmykov:2012rr,delaCruz:2019skx,Klausen:2019hrg,Ananthanarayan:2022ntm}. 
Simple examples of Feynman integrals whose solutions can be expressed in terms of multi-variable hypergeometric functions are shown in Fig.~\ref{fig:exSimplePheynman}. On the left, a massless pentagon
integral is depicted, which can be expressed through the Appell function $F_3$ \cite{Kniehl:2010aj}. On the right is the well-known two-loop sunset integral, which can be expressed through the Appell function $F_2$ \cite{Kniehl:2005bc,Tarasov:2006nk}.
The hypergeometric functions arising in computations of Feynman diagrams generally come with indices  linearly dependent on the parameter of dimensional regularization $\ep$, and one is interested in their Laurent expansions in $\ep$ up to some specified order. There are several computational packages for expansions of different hypergeometric functions, both numerically and analytically \cite{Huber:2005yg,Huber:2007dx,Moch:2005uc,Weinzierl:2002hv,Ablinger:2013cf,Huang:2012qz,Bera:2023pyz,Bezuglov:2023owj}. 

In the present work, we are dealing with the problem of high-precision numerical evaluation of Lauricella functions with indices linearly dependent on some parameter $\ep$ for arbitrary values of their arguments in terms of Laurent series about $\ep = 0$. This problem is ultimately related to the problem of analytic continuation of their defining series representations from their convergence regions to the whole $\mathbb{C}^n$ space, where $n$ is their number of arguments. The known Euler \cite{Exton} and Mellin--Barnes \cite{Exton,AppelKampedeFeriet} integral representations of these functions, in principle, give such analytic continuations. However, from the point of view of both their high-precision numerical evaluations and the general solution of the problem of analytic continuation to the whole $\mathbb{C}^n$ space, it is desirable to have suitable generalized hypergeometric series representations of them in different subdomains covering the whole $\mathbb{C}^n$ space. These series representations should solve  partial differential equations satisfied by these functions and agree in the intersection regions of different subdomains. This is precisely how the analytic continuation is understood in the original works of Kummer and Riemann \cite{AnalyticCont1,AnalyticCont2} and is closely related to the problem of finding the monodromy group. There are mainly two approaches in the literature for analytic continuations of Lauricella functions. One is based on re-expansions of hypergeometric series with the use of known analytic continuations of hypergeometric functions with lower numbers of arguments \cite{AppelKampedeFeriet,Erdelyi,Olsson,Exton,Ananthanarayan:2024nsc,Bera:2024hlq} and the other employs Mellin--Barnes integral representations \cite{hahne1969analytic,Bezrodnykh1,Bezrodnykh2,Bezrodnykh3,Bezrodnykh4,Bezrodnykh5,BezrodnykhReview,HYPERDIRE1,HYPERDIRE2,HYPERDIRE3,HYPERDIRE4}.  

Here, to solve the analytic-continuation problem, we use Frobenius generalized power solutions of the corresponding Pfaffian differential equation systems, satisfied by Lauricella functions, along suitable continuation paths. The series one obtains within this approach are one-dimensional as opposed to multi-dimensional series arising in the approaches based on re-expansions of hypergeometric series or Mellin--Barnes integral representations. The latter property offers us the opportunity to have much more accurate numerical results for Lauricella functions than calculations with multiple sums  \cite{colavecchia2001numerical,colavecchia2004f1,Ananthanarayan:2021bqz,Bera:2024hlq}. To automate the whole computational procedure, we developed the \texttt{PrecisionLauricella} software package written in Wolfram Mathematica language. 

The remainder of the paper is organized as follows. In Section~\ref{sec:LauricellaFunctions}, we briefly recall the definition of Lauricella functions and their treatment in the GKZ and Horn approaches to hypergeometric functions. Section~\ref{sec:NumericalEvaluation} contains all the details on solving the corresponding Pfaffian differential equation systems with Frobenius generalized power series solutions, on their analytic continuation and on the treatment of $\ep$ expansions. Finally, in Section~\ref{sec:Conclusion}, we present our conclusion and discussion of future directions. Appendix~\ref{appendix::PrecisionLauricella} contains details on the usage of the \texttt{PrecisionLauricella} package and its performance. 

\label{key}

\section{Lauricella functions}

\label{sec:LauricellaFunctions}

Lauricella functions \cite{Lauricella} are one of the most natural generalizations of the Gauss hypergeometric function to the case of several variables \cite{bateman1953higher,Exton,FromGaussToPainleve,TheoryHypergeometricFunctions,Schlosser:2013hbz}. Introducing the notation $\bx^\bm = x_1^{m_1}\cdots x_n^{m_n}$, $(\ba)_\bm = (a_1)_{m_1}\cdots (a_n)_{m_n}$ for $n$-tuples of complex numbers $\bx = (x_1,\ldots ,x_n)$, $\ba = (a_1,\ldots ,a_n)$ and of non-negative integers $\bm = (m_1,\ldots ,m_n)$, the four Lauricella series are defined as 
\begin{eqnarray}
F_A^{(n)}(a,\bb,\bc|\bx) &=& \sum_{\bm} \frac{(a)_{|\bm|}(\bb)_{\bm}}{(\bc)_{\bm}\bm !}\bx^\bm\, , \nonumber\\
F_B^{(n)}(\ba, \bb, c| \bx) &=& \sum_{\bm} \frac{(\ba)_\bm (\bb)_\bm}{(c)_{|\bm|}\bm !}\bx^\bm\, , \nonumber\\
F_C^{(n)}(a,b,\bc |\bx) &=& \sum_{\bm} \frac{(a)_{|\bm|}(b)_{|\bm|}}{(\bc)_{\bm} \bm !}\bx^{\bm}\, , \nonumber\\
F_D^{(n)}(a, \bb, c |\bx) &=& \sum_{\bm}\frac{(a)_{|\bm|}(\bb)_{\bm}}{(c)_{|\bm|}\bm!}\bx^\bm\, ,
\end{eqnarray}
where $\bm ! = m_1!\cdots m_n!$, $|\bm| = m_1+\ldots +m_n$, and the summations $\bm$ run over $\mathbb{Z}_{\geq 0}^n$. These series representations converge absolutely in their specific regions. For the $F_B^{(n)}$ and $F_D^{(n)}$ series, the convergence region is given by $|x_i|<1$, $i=1,\ldots n$, while the $F_A^{(n)}$ and $F_C^{(n)}$ series converge in the regions $\sum_{i=1}^n |x_i| < 1$ and $\sum_{i=1}^n |x_i|^{1/2} < 1$, respectively. In what follows, we will focus on the three Lauricella functions $F_A^{(n)}$, $F_B^{(n)}$, and $F_D^{(n)}$ for $n \leq 3$. These functions are selected because, in some cases, they can be expressed in terms of multiple polylogarithms\footnote{See also Ref.~\cite{MPLsHopf1} for general introduction.} (MPLs) \cite{goncharov2,goncharov3}, facilitating independent numerical verification. In the single-variable case, these functions, by definition, reduce to the Gauss hypergeometric function,
\begin{equation}
F_A^{(1)} = F_B^{(1)} = F_D^{(1)} = {}_2F_1\, , 
\end{equation}
and, in the case of two variables, they reduce to Appell functions \cite{AppellFunctions},
\begin{equation}
F_A^{(2)} = F_2\,, \qquad F_B^{(2)} = F_3\,, \qquad F_D^{(2)} = F_1\,.
\end{equation}

We will also allow for the indices of the considered Lauricella functions to depend on some parameter $\ep$ and will we interested in series expansions with respect to the latter. Such expansions may be considered as a model for the evaluation of Feynman integrals as Laurent series with respect to the parameter of dimensional regularization. To evaluate Lauricella functions outside the convergence regions of the corresponding series, we need to perform their analytic continuation. This is most conveniently done with the use of differential equations satisfied by the latter. There are essentially two approaches in the study of multi-variable hypergeometric functions where such differential equations appear. One is based on hypergeometric $\mathcal{A}$-systems of GKZ type \cite{GKZ1,GKZ2,GKZ3,GKZ4,GKZ5} and the other one is Horn's power series approach \cite{Hornapproach}. The general GKZ system\footnote{See Ref.~\cite{GKZhyper}, for a general introduction.} for a function $\Phi$ of $N$ variables $u_1,\ldots ,u_N$ is constructed from a vector $\bc\in \mathbb{C}^{k+1}$ and an $N$-element subset $\mathcal{A} = \{\ba_1,\ldots \ba_N\}\subset \mathbb{Z}^{k+1}$ generating the lattice $\mathbb{Z}^{k+1}$. It is required that there exists a group homomorphism $h:\mathbb{Z}^{k+1}\to \mathbb{Z}$, such that $h(\ba) = 1$ for all $\ba \in \mathcal{A}$. The matrix $\mathcal{A}$ is further associated with the sublattice $\mathbb{L}\subset \mathbb{Z}^{N}$ defined by
\begin{equation}
\mathbb{L} = \{(l_1,\ldots , l_N)\in \mathbb{Z}^N | l_1 \ba_1+\ldots + l_N\ba_N=0\}\,,
\end{equation}
and the GKZ system associated with $\mathcal{A}$ and $\bc$ is given by the following two sets of differential equations. First, we have a system of $k+1$ first-order differential equations
\begin{equation}
\ba_1 u_1\frac{\partial\Phi}{\partial u_1} + \ldots + \ba_N u_N \frac{\partial\Phi}{\partial u_N} = \bc\Phi\,,
\label{eq:GKZ1}
\end{equation}
and, second, an infinite set of equations of the order of $N$ or less,
\begin{equation}
\prod_{l_i<0}\left(
\frac{\partial}{\partial u_i}
\right)^{-l_i} \Phi = \prod_{l_i>0}\left(
\frac{\partial}{\partial u_i}
\right)^{l_i} \Phi\,,
\label{eq:GKZ2}
\end{equation}
each of which corresponds to the vector $(l_1,\ldots ,l_N)\in\mathbb{L}$, where, on the left-hand side of Eq.~\eqref{eq:GKZ2}, the product goes over negative components and the right-hand side involves only positive ones. The solution of Eqs.~\eqref{eq:GKZ1}--\eqref{eq:GKZ2} is given by (formal) power series (so called $\Gamma$-series),
\begin{equation}
\Phi_{\mathbb{L},\underline{\gamma}}(u_1,\ldots ,u_N) = 
\sum_{(l_1,\ldots ,l_N)\in\mathbb{L}}\prod_{j=1}^N\frac{u_j^{\gamma_j+l_j}}{\Gamma (\gamma_j + l_j + 1)}\,,
\label{GammaSeries} 
\end{equation}
where the vector $\underline{\gamma} = (\gamma_1,\ldots ,\gamma_N)$ is connected to the matrix $\mathcal{A}$ and the parameter vector $\bc$ through the relation $\bc = \sum_{j=1}^N \gamma_j \ba_j$. There are well-known relations between Lauricella series and $\Gamma$-series. For example, in the case of the $F_D^{(n)}$ series, we find
\begin{equation}
\frac{(a)_{|\bm|}(\bb)_{\bm}}{(c)_{|\bm|}\bm !} = \Gamma (1-a)\Gamma (c) \prod_{i=1}^n \Gamma (1-b_i)\cdot \prod_{j=1}^N\frac{1}{\Gamma (1+\gamma_j + l_j)}\,,
\end{equation}
with $N=2n+2$, $\underline{\gamma} = (\gamma_1,\ldots ,\gamma_N) = (c-1,-b_1,\ldots ,-b_n, -a, 0,\ldots , 0)$ and 
\begin{equation}
(l_1,\ldots , l_N) = (m_1,\ldots ,m_n)\cdot \left(\bf{1}_n, -\mathbb{I}_n, -\bf{1}_n, \mathbb{I}_n \right)\, ,
\end{equation}
where {$\bf{1}_n$} is the column vector and $\mathbb{I}_n$ is the $n\times n$ identity matrix. So, for the $\mathbb{L}$ lattice,\footnote{For the form of the corresponding $\mathcal{A}$ matrix, see for example Refs.~\cite{BezrodnykhReview,GKZhyper}.} we take the lattice spanned by the rows of the above $n\times N$ matrix.  Now, substituting $u_j  = 1$ for $1\leq j\leq n+2$ and $u_j = x_{j-n-2}$ for $n+3\leq j\leq 2n+2$, the corresponding $\Gamma$-series turn into a Lauricella power series,
\begin{equation}
\Phi_{\mathbb{L},\underline{\gamma}}(1,\ldots ,1,z_1,\ldots ,z_n) = \left(\prod_{j=1}^{n+2}\Gamma (1+\gamma_j)^{-1}\right) F_D^{(n)}(a,\bb,c|\bx)\, .
\end{equation}

Within Horn's power series approach \cite{Hornapproach}, one considers general series of the form 
\begin{equation}
F(x_1,\ldots, x_n) = \sum_{\bk\in \mathbb{Z}^n} \Lambda_{k_1,\ldots,k_n} x_1^{k_1}\dots x_n^{k_n} = \sum_{\bk\in \mathbb{Z}^n}\Lambda (\bk) \bx^\bk\,.
\end{equation}
It is said that this series is of hypergeometric type if the ratio of any two adjacent coefficients is a rational function of $\bk$ vector components. That is,  we have
\begin{equation}
\frac{\Lambda (\bk + \be_i)}{\Lambda (\bk)} = \frac{P_i (\bk)}{Q_i(\bk)}\, , \label{eq:HornRatio}
\end{equation}
with $i=1,\ldots n$, where $P_i(\bk)$ and $Q_i(\bk)$ are some polynomials in $n$ variables and $\be_i$ is the basis vector with the $i$-th component equal to one and all others equal to zero. From here, it is easy to see that the function $f(\bx)$ solves the following system of partial differential equations \cite{Hornapproach,AppelKampedeFeriet,Mellin}: 
\begin{equation}
Q_i (\theta) (x_i^{-1} F(\bx)) = P_i (\theta) F(\bx)\, ,
\end{equation}
where the differential operators $P_i(\theta)$ and $Q_i(\theta)$ are obtained by substituting their $\bk$ vector arguments with vectors of derivatives $\theta = (\theta_1,\ldots ,\theta_n)$ with $\theta_i = x_i\partial/\partial x_i$.   For example, the $F_D^{(n)}$ Lauricella series with
\begin{equation}
\Lambda (\bk) = \frac{(a)_{|\bk|}(\bb)_{\bk}}{(c)_{|\bk|}\bk !}\,,
\end{equation}
satisfies Eq.~\eqref{eq:HornRatio} with 
\begin{equation}
P_i (\bk) = (a+|\bk|)(b_i+k_i)\, ,\qquad Q_i(\bk) = (c+|\bk|)(1+k_i)\, ,
\end{equation}
where $|\bk| = \sum_{j=1}^n$,
and the corresponding system of partial differential equations takes the form
\begin{equation}
(c+\sum_{m=1}^n\theta_m)(1+\theta_i)(x_i^{-1} f(\bx)) = (a+\sum_{m=1}^n\theta_m)(b_i+\theta_i)f(\bx)\, .
\end{equation}
with $i=1,\ldots ,n$.
Similarly, in the case of the Appell function $F_D^{(2)}(\alpha; \beta_1, \beta_2; \gamma; x, y) = F_1$, the  corresponding system of partial differential equations is given by
\begin{eqnarray}
\left[x(1-x)\frac{\partial^2}{\partial x^2} + y(1-x)\frac{\partial^2}{\partial x \partial y} + \left(\gamma - (\alpha + \beta_1 + 1)x\right)\frac{\partial}{\partial x} - \beta_1 y \frac{\partial}{\partial y} - \alpha \beta_1\right] F_1 & =& 0\,,
\nonumber \\
\left[y(1-y)\frac{\partial^2}{\partial y^2} + x(1-y)\frac{\partial^2}{\partial x \partial y} + \left(\gamma - (\alpha + \beta_2 + 1)y\right)\frac{\partial}{\partial y} - \beta_2 x \frac{\partial}{\partial x} - \alpha \beta_2\right] F_1 & =& 0\,.\quad
\end{eqnarray}

\section{Numerical evaluation}
\label{sec:NumericalEvaluation}

Having obtained the partial differential equations satisfied by Lauricella functions, let us proceed with their high-precision numerical solution. The latter is most conveniently done  with the use of the well-known Frobenius method \cite{coddington1984theory,CattaniLectures,SaitoSturmfelsTakayama,Dmodulesholonomicfunctions,Koutschan2010}.\footnote{For the application of the Frobenius method in the context of Feynman diagrams, see for example Refs.~\cite{Frobenius1,Frobenius2,Frobenius3,Frobenius4,Frobenius5,KKOVelliptic2,Moriello:2019yhu,Bonciani:2019jyb,Frellesvig:2019byn,DiffExp,Bonisch:2021yfw, Bezuglov:2021tax, Blumlein:2021hbq,Bezuglov:2022npo,Armadillo:2022ugh}.} While obtaining systems of partial differential equations is straightforward, applying the Frobenius method to them can be challenging. First, one needs to transform the available partial differential equations to the so-called Pfaffian form,
\begin{equation}
dJ = M J = \left(M_1 \, dx_1 + M_2 \, dx_2 + \dots + M_n \, dx_n\right) J\,, \label{eq:PfaffianForm}
\end{equation}
for some basis of functions $J$. These systems can be derived through various methods, among which the most advanced ones are based on Gr\"obner basis techniques for D-modules.\footnote{We refer the interested reader to Refs.~\cite{SaitoSturmfelsTakayama,Dmodulesholonomicfunctions,Koutschan2010,Chestnov:2022alh,Henn:2023tbo}.} In our work, we utilize pre-derived Pfaffian systems from Ref.~\cite{Bezuglov:2023owj} with the following function bases:
\begin{eqnarray}
&&\left\{\theta_{x_{j_1}} \dots \theta_{x_{j_k}} F_N^{(n)} \,\Big|\, 0 \leq k \leq n,\, j_1 < j_2 < \dots < j_k \right\}\,, \qquad N = A, B\,,
\nonumber\\
&&\left\{F_D^{(n)},\, \theta_{x_{j}} F_D^{(n)} \,\Big|\, j = 1, \dots, n \right\}\,.
\end{eqnarray}

Our aim is to evaluate the value of a chosen Lauricella function at the point $\bx$ given its initial value at the point $\bx_0$. This is most easily done by considering the above differential equation system along some path connecting these two points. Suppose we want to get a solution along the path $f$  parameterized by the parameter $t$, such that 
\begin{equation}
x_1 = f_1(t)\,,\quad x_2 = f_2(t)\,,\quad \dots\,,\quad x_i = f_i(t)\,.
\end{equation}
In this way, the differential equation system in Eq.~\eqref{eq:PfaffianForm} restricted to this path takes the form
\begin{equation}
\frac{dJ}{dt} = M_t J\,, \qquad M_t = \frac{\partial x_1}{\partial t}M_1+\frac{\partial x_2}{\partial t}M_2+\dots+\frac{\partial x_n}{\partial t}M_n\,.  \label{eq:Mteqn}
\end{equation}
In practice, it is usually convenient to choose the path $f$ as a line passing through the origin, where we apply our boundary conditions,
\begin{equation}
  x_1 = \kappa_1 t\,,\quad x_2 = \kappa_2 t\,,\quad\dots\,,\quad x_n = \kappa_{n} t\,,
\label{eq:fpath}
\end{equation}
where $\kappa_i\in\mathbb{R}_{\geq 0}$ with $i=1,\ldots,n$.
This will allow us to reach many points in $\mathbb{C}^n$ without crossing the fixed cuts\footnote{For complex $\kappa_i$ , the branch cuts of Riemann sheets for Lauricella function will generally rotate after projection onto the t-plane through $x_i \rightarrow \kappa_i t$.} in the variable $t$. For the definition of cuts and the modification of the path for the residual points, see Section~\ref{sec:AnalyticalContinuation} on analytic continuation. Note also that, for example, for points with $x_1=0$, it is more natural to consider the initial Pfaffian differential equation system in lower, $n-1$ dimensions.

For example, for the Appell $F_1$ function, the initial Pfaffian system of differential equations has the form
\begin{equation}
dJ = \left( M_x dx + M_y dy\right) J\,,
\end{equation}
where
\begin{eqnarray}
M_x &=& \left(
\begin{array}{ccc}
0 & \frac{1}{x} & 0 \\
-\frac{\alpha  \beta_1}{x-1} &
\frac{-\left(x^2 (\alpha +\beta_1)\right)+x (\gamma +y (\alpha
	+\beta_1-\beta_2)-1)+y (\beta_2-\gamma
	+1)}{(x-1) x (x-y)} &
\frac{\beta_1 (y-1)}{(x-1)
	(x-y)} \\
0 & \frac{\beta_2 y}{x^2-x y} &
-\frac{\beta_1}{x-y} \\
\end{array}
\right)\,,
\nonumber\\
M_y &=& \left(
\begin{array}{ccc}
0 & 0 & \frac{1}{y} \\
0 & \frac{\beta_2}{x-y} &
-\frac{\beta_1 x}{y (x-y)} \\
-\frac{\alpha \beta_2}{y-1} &
-\frac{\beta_2 (x-1)}{(y-1)
	(x-y)} & \frac{y (-\gamma +y (\alpha
	+\beta_2)+1)-x (\beta_1-\gamma +y (\alpha -\beta_1+\beta_2)+1)}{(y-1) y
	(x-y)} \\
\end{array}
\right)\,,
\end{eqnarray}
and the vector of basis functions is given by
\begin{equation}
J_1  = \left\{F_1, x\frac{\partial}{\partial x}F_1, y\frac{\partial}{\partial y}F_1 \right\}\,. 
\end{equation}
Now, choosing the path as
\begin{equation}
x = t\,, \quad y = \kappa t\,,
\end{equation}
with $\kappa\in\mathbb{R}$, the corresponding $M_t$ matrix takes the form
\begin{equation}
M_t = \left(
\begin{array}{ccc}
0 & \frac{1}{t} & \frac{1}{t} \\
-\frac{\alpha  \beta_1}{t-1} &
-\frac{-\gamma +\alpha  t+\beta_1 t+1}{(t-1) t} & -\frac{\beta_1}{t-1} \\
-\frac{\alpha  \beta_2 \kappa
}{\kappa  t-1} & -\frac{\beta_2 \kappa }{\kappa  t-1} &
-\frac{-\gamma +\alpha  \kappa 
	t+\beta_2 \kappa  t+1}{t
	(\kappa  t-1)} \\
\end{array}
\right)\,.
\end{equation}
Having obtained the $M_t$ differential equation system, the next step is its solution.

\subsection{Series solutions at regular and singular points}
\label{sec:SeriesSolution}

Let us first consider the solution of the $M_t$ differential-equation system in Eq.~\eqref{eq:Mteqn} in the neighborhood of a singular point. Without loss of generality, we may assume that this point is given by $t=0$. Moreover, we will assume\footnote{For hypergeometric functions, this is always the case.} that the system is Fuchsian at $t=0$. The $M_t$ matrix for such a system can be written as
\begin{equation}
M_t = \frac{A_0}{t} + R(t)\, ,
\end{equation}
where $R(t)$ is some rational function in the variable $t$. Its exact form is not particularly important, as long as it does not have poles at $t = 0$.\footnote{Here, we focus on the solution in the vicinity of $t = 0$, and its further analytic continuation to other regions will be considered in the next subsection.} For Lauricella functions with indices that are linearly dependent on some parameter $\ep$, the matrices $A_0$ and $R(t)$  will also depend rationally on  $\ep$. Besides that, no other restrictions are imposed on these matrices. Note also that the expansion about the regular point $t=0$ corresponds to $A_0 = 0$ with all technicalities presented below remaining valid.

The fundamental series solutions matrix $U$ for system in Eq.~\eqref{eq:Mteqn} satisfies the following set of equations:
\begin{equation}
\frac{d U}{dt}=M_t U\,, \qquad \text{det}\left(U\right)\ne 0\, ,
\end{equation}
which means that the columns of this matrix represent linearly independent solutions of Eq.~\eqref{eq:Mteqn}. Any particular solution can be obtained from the matrix $U$ by multiplying it with a vector of boundary conditions. We will look for a generalized power series expansion of this matrix in the form\footnote{See Refs.~\cite{Frobenius4,Frobenius5}, for a similar consideration.} 
\begin{equation}
U =\sum\limits_{\lambda \in S}t^{\lambda}U^{(\lambda)}= \sum\limits_{\lambda \in S}t^{\lambda}\sum\limits_{n = 0}^{\infty}\sum_{k = 0}^{m_{\lambda}}c^{(\lambda)}_{n,k} t^n \log^k(t)\,,
\label{eq:FrobeniusAn}
\end{equation}
where the set $S$ consists of the non-resonant eigenvalues of the $A_0$ matrix and $m_{\lambda}$ is the number of eigenvalues in resonance with eigenvalue $\lambda$, that is their differences are integers.  
If several eigenvalues are in resonance, then the set $S$ will only include the smallest of them. For example, suppose the set of eigenvalues of the matrix $A_0$ is given by
$\{0,-1,-2, 1-\ep,1-2\ep,-2\ep,1/2-\ep,$  $1/2-\ep\}$.  Then $S = \{-2,1-\ep,-2\ep,1/2-\ep\}$,  
$m_{-2} = 2$, $m_{1-\ep} = 0$, $m_{-2\ep} = 1$ and $m_{1/2-\ep} = 1$. 

The recurrence relations for the coefficients $c^{(\lambda)}_{n,k}$ can be found by  substituting the anzatz of Eq.~\eqref{eq:FrobeniusAn} into differential equation system in Eq.~\eqref{eq:Mteqn}. Having the latter, we can solve them numerically up to any predetermined order $n$ and have a numerical solution of the fundamental system with any desired accuracy. However, in general, the recurrence relations obtained in this way will not have finite order. To avoid that, let us first multiply both sides of our differential system with a polynomial matrix $Q(t)$, such that $tQM_t$ also becomes polynomial in $t$. Although this condition restricts the form of $Q$, its choice remains somewhat arbitrary. We constrain it further by requiring that $Q$ is diagonal, its degree in $t$ is minimal and  $Q\Big|_{t=0} = 1$. Due to the presence of resonance eigenvalues in $A_0$, we cannot write a general formula expressing the new coefficients in terms of already determined ones. For that reason, we introduce an $l$-shift parameter which depends on $\lambda$ and determines different modes for the solution of recurrence relations. We define this parameter as
\begin{equation}
l_{\lambda} = \text{max}( s \cap \mathbb{Z})\,,
\label{eq:lShiftFrob}
\end{equation}
where $s$ is the set of solutions to the polynomial equation $\det\left(A_0 - (n+\lambda)I\right) = 0$ with respect to $n$. Since $\lambda$ is selected from the set of $A_0$ eigenvalues $S$, $l_{\lambda}$ is always greater than zero. Having done that, the recurrence relations for the matrix coefficients $c^{\lambda}(n,k)$ can be written as
\begin{eqnarray}
&&\hspace{-1cm}c_{n,k}^{(\lambda)} = -\left(A_0 - (n+\lambda)I\right)^{-1}\left[\hat{M}_1c_{n,k}^{(\lambda)} - \hat{Q}_1(n+\lambda)c_{n,k}^{(\lambda)} - (k+1)\hat{Q}c_{n,k+1}^{(\lambda)}\right]\,, \qquad n > l_{\lambda}\,,
\qquad\label{eq:FrobGeneralEq}
\\
&&\hspace{-1cm}\hat{M}_1c_{n,k}^{(\lambda)} - \hat{Q}_1(n+\lambda)c_{n,k}^{(\lambda)} - (k+1)\hat{Q}c_{n,k+1}^{(\lambda)}+\left(A_0 - (n+\lambda)I\right)c_{n,k}^{(\lambda)}  = 0\,, \qquad 0 \le n \le l_{\lambda}\,,
\qquad\label{eq:FrobBoundaryEq}
\end{eqnarray}
with the condition
\begin{equation}
c_{n,k}^{(\lambda)} = 0\,,
\label{eq:CBoundaryes}
\end{equation}
for $n<0$ or $k > m_{\lambda}$.
Here, $M_1 = (tQM_t - A_0) $, $Q_1 = Q-1$, and we have introduced the $\hat{}$ operator as
\begin{equation}
\hat{H}c_n = \sum\limits_{i=0}^n c_ih_{n-i}\,,
\label{eq:hatOperator}
\end{equation}
where $h_n$ are the coefficients in the power series $H = \sum\limits_{n = 0}^{\infty}h_nt^n$.
From the definitions of the matrices $Q_1$ and $M_1$, it is evident that the right-hand side of Eq.~\eqref{eq:FrobGeneralEq} only contains coefficients $c_{m,k}^{(\lambda)}$ with $m < n$. It is also clear that the number of non-trivial equations in Eq.~\eqref{eq:FrobBoundaryEq} is less than the number of unknowns. Therefore, the solution will depend on a set of parameters. These parameters can be defined arbitrarily, with the only condition that $\det U \ne 0$. They may either be matched with the matrix $t^{A_0}$ or generated randomly.

Altogether, to obtain a Frobenius series solution at the point $t=t_p$ (within the convergence region of the series) from the boundary condition vector $J_0$ at $t=0$ with the prescribed numerical accuracy and given expansion order in the parameter $\ep$, we should perform the following steps:

\begin{enumerate}
	\item First of all, we need to define a set $S$ of non-resonant eigenvalues over which the summation will be carried out. To do this, we obtain the matrix $A_0$ and calculate its eigenvalues. If several eigenvalues are in resonance, then the set $S$ will include only the smallest one of them. To each element of $S$, we associate an integer $m_{\lambda}$, the number of eigenvalues in resonance with eigenvalue $\lambda$. This number only determines the upper limit of the possible power of the logarithm for a given eigenvalue in Eq.~\eqref{eq:FrobeniusAn}, and the maximum power in the obtained solution may be lower.
	
	\item For each element from the set $S$, we determine the $l$-shift from Eq.~\eqref{eq:lShiftFrob}.
	
	\item Next, we determine the matrix $Q$. We define it as $Q = \text{diag}\left(Q_1,\dots,Q_n\right)$, where $Q_i$ is the least common multiple of the set of denominators of the $i$-th row of the matrix $tM_t$, and normalize it as $Q\Big|_{t=0} = 1$.
	 
	\item Now, we need to define the matrix of boundary conditions $U_0$. To do this, for each eigenvalue $\lambda \in S$, it is necessary to solve the system of linear equations in Eq.~\eqref{eq:FrobBoundaryEq}. These equations are written for $0\le k \le m_{\lambda}$, and we set $c_{n,k}^{(\lambda)} = 0$ if $n < 0$. Then, the matrix of boundary conditions can be written as $U_0 =\sum_{\lambda \in S}t^{\lambda}U^{(\lambda)}_0= \sum_{\lambda \in S}t^{\lambda}\sum_{n = 0}^{l_{\lambda}}\sum_{k = 0}^{m_{\lambda}}c_{n,k}^{(\lambda)} t^n \log^k(t)$.
	As mentioned earlier, the choice of such a matrix is somewhat arbitrary, except that we should have $\det U_0 \ne 0$. We also require that this matrix does not contain poles in the parameter $\ep$. 
	
	\item Using the set of $U_0^{(\lambda)}$ matrices, one can determine the actual maximum powers of the logarithms $m_{\lambda}$ in ansatz~\eqref{eq:FrobeniusAn} to be used in the next step for the full solution.
	
	\item Now, we simply solve system in Eq.~\eqref{eq:FrobGeneralEq} numerically, first expanding it into a series in $\ep$. The boundary conditions in this case are obtained from the $U_0$ matrix. Having found solutions for the required $c_{n,k}^{(\lambda)}$ coefficients,  we get the solution for the $U$ matrix at the point $t_p$. We stop the summation when the next term in the obtained series is less than the required accuracy by several orders of magnitude.
	
	\item  The final solution is given by $U(U_0)^{-1}J_0$.
\end{enumerate}

As a simple example, we consider the function ${}_2 F_1 \left(\begin{array}{c}1/2 + 2\ep, 1/2 \\ 2 \end{array} \Big|t\right)$. The differential equation for this function can be easily obtained, and the corresponding $M_t$ and $A_0$ matrices are given by
\begin{equation}
M_t =\left(
\begin{array}{cc}
0 & \frac{1}{t} \\
\frac{2 \varepsilon +\frac{1}{2}}{2 (1-t)} & \frac{2
	\varepsilon  t+t-1}{(1-t) t} \\
\end{array}
\right)\,, \qquad
A_0 =\left(
\begin{array}{cc}
0 & 1 \\
0 & -1 \\
\end{array}
\right)\,,
\end{equation}
together with the boundary conditions vector $J_0 = \{1,0\}^{\top}$. The eigenvalues of the $A_0$ matrix are $\{-1,0\}$, and we conclude that the set $S$ will consist of only one element $S = \{-1\}$. Also, we have $l_{-1} = 1$ and $m_{-1} = 1$. For the $Q$ matrix, we can take $Q = \text{diag}\left(1,1-t\right)$.

The matrix of boundary conditions $U_0$ is then calculated to be
\begin{equation}
U_0 =\left(
\begin{array}{cc}
-\varepsilon  \log (-t)-\frac{1}{t}+\frac{\log
	(-t)}{4} & 1 \\
-\varepsilon +\frac{1}{t}+\frac{1}{4} & 0 \\
\end{array}
\right)\,,
\end{equation}
from where we get $C_0 = \{0,1\}^{\top}$, so that $U_0 C_0 = J_0$.
The system of recurrence relations for the $c_{n,k}^{(\lambda)}$ coefficients in Eq.~\eqref{eq:FrobGeneralEq} in this case is given by  
\begin{eqnarray}
c_{n,1}^{(-1)} &=& \left(
\begin{array}{cc}
\frac{4 \varepsilon +1}{4 (n-1) n} & \frac{2
	\varepsilon +n-1}{(n-1) n} \\
\frac{4 \varepsilon +1}{4 n} & \frac{2 \varepsilon
	+n-1}{n} \\
\end{array}
\right)c_{n-1,1}^{(-1)}\, ,
\nonumber\\
c_{n,0}^{(-1)} &=& \left(
\begin{array}{cc}
-\frac{1}{n-1} & -\frac{1}{(n-1) n} \\
0 & -\frac{1}{n} \\
\end{array}
\right)c_{n,1}^{(-1)} + \left(
\begin{array}{cc}
0 & \frac{1}{(n-1) n} \\
0 & \frac{1}{n} \\
\end{array}
\right)c_{n-1,1}^{(-1)}
+
\left(
\begin{array}{cc}
\frac{4 \varepsilon +1}{4 (n-1) n} & \frac{2
	\varepsilon +n-1}{(n-1) n} \\
\frac{4 \varepsilon +1}{4 n} & \frac{2 \varepsilon
	+n-1}{n} \\
\end{array}
\right)c_{n-1,0}^{(-1)}\, ,
\end{eqnarray}
which is valid for  $n > l_{-1} = 1$. This system can be easily solved numerically and, for example, at the point $t_p = 1/2$, we get
\begin{equation}
{}_2 F_1 \left(\begin{array}{c}1/2 + 2\ep, 1/2 \\ 2 \end{array} \Big|\frac{1}{2}\right) = 1.0787052023767587133\dots +( 0.34115988312544546717\dots) \ep + \OO(\ep^2)\,.
\end{equation}

\subsection{Analytic continuation}
\label{sec:AnalyticalContinuation}

The Frobenius power series solutions obtained in the previous subsection are valid only in their regions of convergence, which are given by the distances to their nearest-neighbor singularities.\footnote{Sometimes, the convergence region can be improved, for example, through the use of a M\"obius transformation, which repositions the nearest singularities \cite{Frobenius4,DiffExp}.} To obtain a solution valid at any point of the complex plane, it is necessary to perform a procedure of analytic continuation. Suppose we want to continue our fundamental series solutions matrix $U(t)$ expanded at $t=t_1$ to another region, which is the convergence region of another fundamental series solutions matrix $U'(t)$ obtained by the expansion at some other regular or singular point $t=t_2$. Assume also that these two regions have non-zero intersection. Then, since these two matrices determine solutions of the same system of differential equations in the intersection region, they should be related by some constant matrix $L$,
\begin{equation}
  U(t) = U'(t)L\,, \qquad \det L \ne 0\,,
\end{equation}  
due to some arbitrariness in the definition of the fundamental series solutions matrix. The $L$ matrix is then easily determined from the values of the two fundamental series solutions matrices at some matching point $t=t_m$ in the intersection region, and we have
\begin{equation}
U(t) = U'(t)\left(U'(t_m)\right)^{-1}U(t_m)\,,
\end{equation}
which is valid in the convergence region of the $U'(t)$ matrix. This procedure can be continued until we determine the fundamental series solutions matrix in the whole complex plane. In the case when the fundamental series solutions matrix contains logarithms and non-integer powers, there are, however, some subtleties. The presence of logarithms and non-integer powers makes our matrix function multi-valued. It can be made single-valued by introducing cuts in the complex plane and so specifying the Riemann sheet where this function is evaluated. It is natural to introduce for logarithms as well as for non-integer powers standard branch cuts, that is horizontal lines parallel to the real axis that go from minus infinity to the singular points. The introduction of cuts has consequences. First, they restrict the regions of validity of our generalized power series. If the convergence region of the generalized power series crosses the cut, then only the region before the cut, containing the expansion point, will give us a series solution  on a Riemann sheet defined by the cuts. Going through the cut takes us to another Riemann sheet. Second, the path we should follow to analytically continue our solution should avoid the introduced cuts.     

Lauricella functions are also multi-valued functions in the variables $x_1,\ldots, x_n$. To make them single-valued, we specify their defining Riemann sheets by introducing the cuts from the points of their singular loci to the point at infinity $(-\infty,\ldots,-\infty)$. These cuts are easily reproduced when considering Lauricella Pfaffian differential equation systems along path~\eqref{eq:fpath} provided that, in the $t$ plane, we have cuts going from $-\infty$ to the singular points parallel to the real axis. We consider such cuts to be fixed in the $t$ plane, that is they always go parallel to real axis and directed towards decreasing values of the real component in the complex plane. This then restricts the parameters $\kappa_i$ in Eq.~\eqref{eq:fpath} and thus the points in $\mathbb{C}^n$ that can be reached with such paths. These are points with  $\Im(x_i) = 0$ and $\Re(x_i)>0$, $\Im(x_i) = 0$ and $\Re(x_i)<0$, $\Re(x_i) = 0$ and $\Im(x_i)>0$ or $\Re(x_i) = 0$ and $\Im(x_i)<0$ simultaneously for all $i=1,\ldots ,n$. To reach other points, we proceed in the following steps. First, starting from the origin, we construct a path to the point $\bx^{(1)}$ obtained from the point $\bx$ by setting to zero the imaginary parts of its components, $\Im (x_i)\to 0$, and also the real parts if they are negative, $\Re (x_i)\to 0$ if $\Re (x_i) < 0$. The path from the origin to the point $\bx^{(1)}$ is then constructed with all $\kappa_i\geq 0$, and the evolution in the $M_t$ system runs in the positive real direction. Then, starting from the point $\bx^{(1)}$ (considered as an origin in the corresponding system of differential equations), we construct a path to the point $\bx^{(2)}$ obtained from the point $\bx$ by setting to zero the imaginary parts of its components, $\Im (x_i)\to 0$.  The evolution in the $M_t$ system this time runs in the negative real direction. Next, staring from the point $\bx^{(2)}$ (again considered as an origin in the corresponding system of differential equations), we construct a path to the point $\bx^{(3)}$ obtained from the point $\bx$ by setting to zero the negative imaginary parts of its components, $\Im (x_i)\to 0$ if $\Im (x_i) < 0$. The evolution in the $M_t$ system now runs in the positive imaginary direction. Finally, starting from the point $\bx^{(3)}$ (again considered as an origin in the corresponding system of differential equations), we construct a path to our point $\bx$.  The evolution in this $M_t$ system will run in the negative imaginary direction. In all steps, the corresponding paths will have  $\kappa_i\geq 0$ for all $i=1,\ldots ,n$ and allow us to eventually  reach all the points in $\mathbb{C}^n$. In addition, we would like to note that, for Lauricella functions, all the singular points will lie on the real axis, and the evolution in the $M_t$ system along the real axis may end up on the $t$ plane cuts. In this case, we need to specify on which side of the cut we evaluate our fundamental matrix system. Here, we follow the $+i \delta$ convention with $\delta > 0$, which is natural as the boundary conditions at $t=0$ are also supplied with this prescription. That is, we assume that, for points on cuts, the fundamental series solutions matrix $U(t)$ is evaluated on their upper sides.  

\begin{figure}[h]
\centering
\begin{minipage}{0.45\textwidth}
\includegraphics[width=\textwidth]{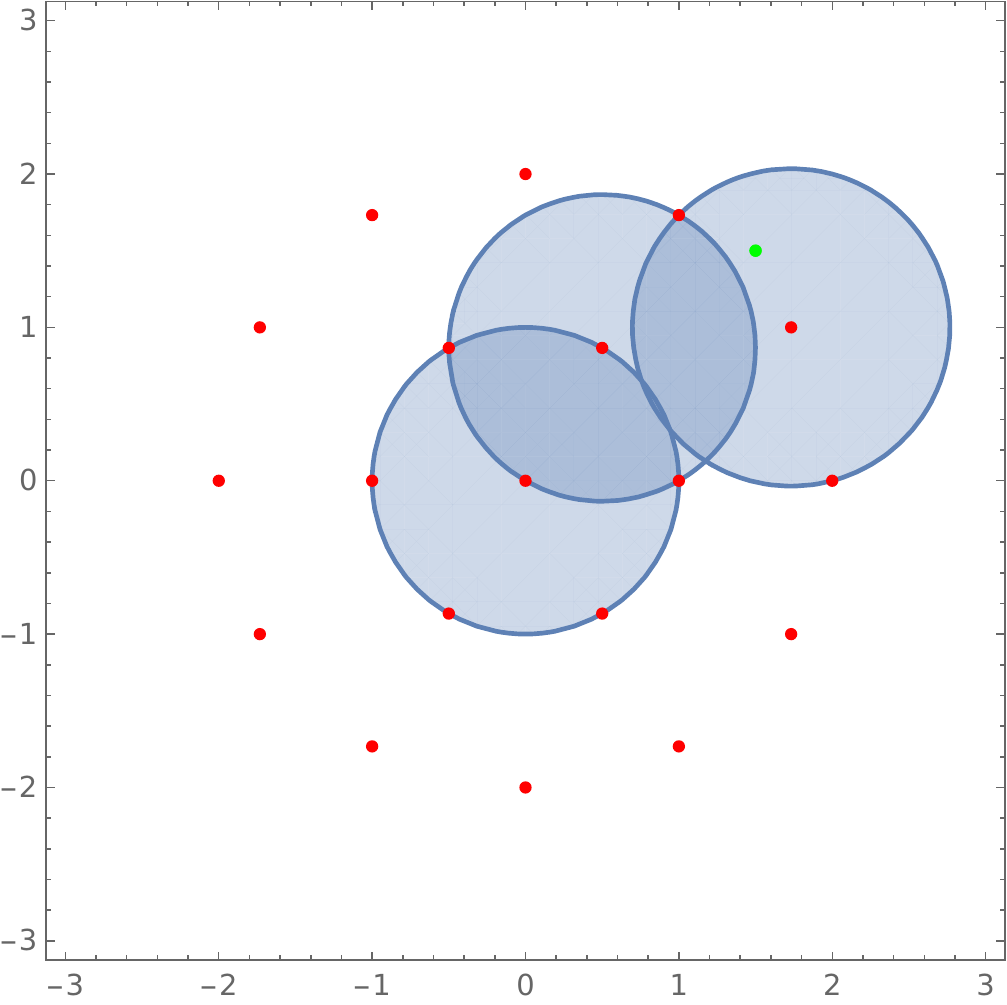}
\label{graph1}
\end{minipage}\hfill
\begin{minipage}{0.45\textwidth}
\includegraphics[width=\textwidth]{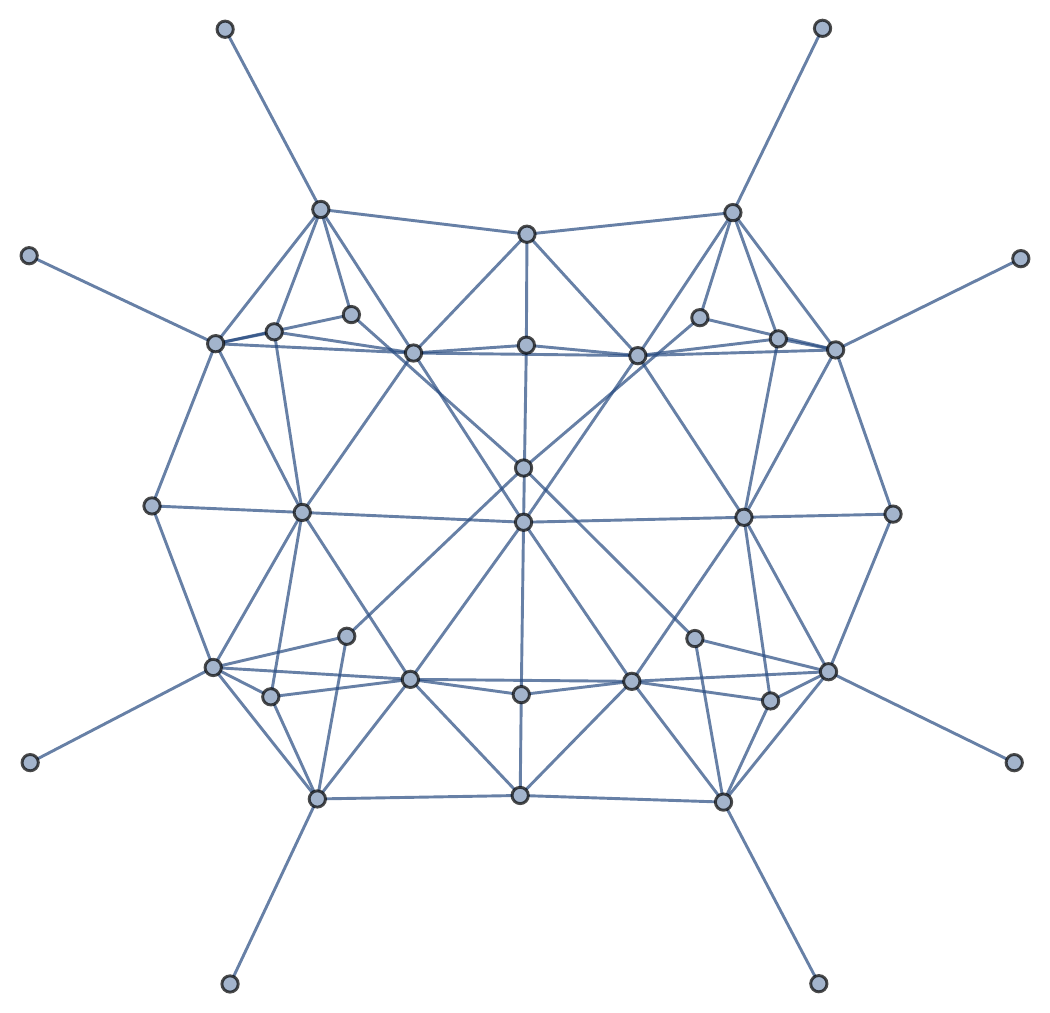}
\label{graph2}
\end{minipage}
\caption{On the left, one of the possible paths with complex singular points is shown. The singular points are shown in red, the end point in green, the starting point is at the origin, and shaded circles represent convergence regions. On the right, the intersection graph of the expansion regions is displayed. The nodes in the intersection graph include expansion regions both at regular and singular points.}
\label{fig:analExample}
\end{figure}

The overall procedure for the analytic continuation of the fundamental series solutions matrix $U(t)$ proceeds in the following steps.  First, in the rectangular area containing the starting and end points, we generate a set of distinct circular regions whose centers lie on a grid. The radii of the regions are given by the distances of their origins to the nearest singularities. Having done that, we compute their intersection graph, which is defined as follows. Each region corresponds to a graph node, and nodes representing regions with starting and end points are marked as special.  Any two nodes are considered to be connected  with an edge if the minimal distance of the points from one region to the origin of the other region is less than three quarters of the radius of the other region.\footnote{This choice is somewhat arbitrary, but we found it to be close to optimal.} The positions of branch cuts are also taken into account. If two regions lie on opposite sides of a branch cut, they are considered as non-intersecting. Knowing the intersection graph, the problem of finding the analytic-continuation path connecting the starting and end nodes is reduced to the problem of finding such a path between the starting and end nodes in a graph. For that purpose, one can use, for example, a built-in Wolfram Mathematica function. The latter generally returns multiple paths, among which, for analytic continuation, we choose the shortest. If there is no path found, we increase the grid step and enlarge the area where the expansion regions are generated. This process is repeated until at least one path is found. However, under normal conditions, the initially generated set of regions is usually sufficient. A simplified example of this procedure is shown in Fig.~\ref{fig:analExample}. 

\subsection{$\varepsilon$ expansion and accuracy estimation}
\label{sec:EpExpansion}

Let us now discuss the question of the $\ep$ expansions of the obtained series solutions. We ignored this question before deliberately, as it does not change the whole consideration and adds only technical details related to the performance of our algorithmic procedure.

The first natural way to solve this problem is through "brute force," that is by considering the series expansions of the coefficients,
\begin{equation}
\label{eq:epExpansionForC}
c_{n,k}^{(\lambda)} = \frac{1}{\ep^p}\sum\limits_{t=0}^{\infty}c^{(\lambda , t)}_{n,k}\ep^t\,,
\end{equation}
truncated beyond the required order in $\ep$. Here, the parameter $p$ stands for the order of the leading singularity in $\ep$ and is determined from the boundary conditions. Substituting this ansatz into Eqs.~\eqref{eq:FrobGeneralEq} and \eqref{eq:FrobBoundaryEq} and gathering terms with similar powers of $\ep$, we immediately get recurrence relations for the expansion coefficients  $c^{(\lambda , t)}_{n,k}$. However, this approach has several drawbacks.  One of them can be illustrated by a simple example. Consider the recurrence relation:
\begin{equation}
c_n= \frac{c_{n-1}}{(n-\ep ) \left(\ep
	+n-\frac{1}{2}\right)}\,.
\label{eq:BruteEx}
\end{equation}
Expanding $c_n$ as a series in $\ep$ and substituting it back into
this equation, we obtain for the first four terms in the series the following system of recurrence relations:
\begin{eqnarray}
c_n^{(0)}&=& -\frac{2 c_{n-1}^{(0)}}{n-2 n^2}\,,
\label{eq:BruteEx1}
\\
c_n^{(1)}&=& -\frac{2
	c_{n-1}^{(0)}}{(1-2 n)^2 n^2}-\frac{2 c_{n-1}^{(1)}}{n-2 n^2}\,,
\label{eq:BruteEx2}
\\
c_n^{(2)}&=&
\frac{\left(8 n^2-4 n+2\right) c_{n-1}^{(0)}}{n^3 (2
	n-1)^3}-\frac{2 c_{n-1}^{(1)}}{(1-2 n)^2 n^2}-\frac{2 c_{n-1}^{(2)}}{n-2
	n^2}\,,
\label{eq:BruteEx3}
\\
c_n^{(3)}&=& -\frac{2 \left(8 n^2-4 n+1\right)
	c_{n-1}^{(0)}}{(1-2 n)^4 n^4}+\frac{\left(8 n^2-4 n+2\right)
	c_{n-1}^{(1)}}{n^3 (2 n-1)^3}-\frac{2 c_{n-1}^{(2)}}{(1-2 n)^2 n^2}-\frac{2
	c_{n-1}^{(3)}}{n-2 n^2}\,.
\label{eq:BruteEx4}
\end{eqnarray}
We see that, although the original recurrence relation was simple, the recurrence relations for the expansion coefficients in $\ep$ start growing in size with the expansion order. There are also less obvious complications. For example, the fundamental matrix of series solutions $U$ and its inverse $U^{-1}$, which arise at different steps of our calculation procedure, may contain poles in $\ep$. While these poles arise at intermediate steps and do not show up in the final answer, processing them requires additional computational resources. Finally, equations like Eqs.~\eqref{eq:BruteEx1}--\eqref{eq:BruteEx4} are poorly suited for parallel computation.

To address these issues, we take a different approach. Instead of keeping $\ep$ as a symbolic parameter, we substitute it with its specific numerical values. These values are chosen as $2n$ lattice points $\ep \in \{\pm m h ~ |~ m = 1,...,n \}$, where $h$ is some small lattice step. The series solutions for each lattice point are then calculated independently. To reconstruct the analytical dependence of the final answer on $\ep$, we use interpolation polynomials. As an example, let us consider the reconstruction of some function $f(\ep)$ not singular in $\ep$. Suppose that we choose the four lattice points 
$\{f(-2h), f(-h), f(h), f(2h)\}$. Using Lagrange interpolation polynomials, we can approximate the series expansion of $f(\ep)$ as 
\begin{eqnarray}
f(\ep)&\approx&\frac{1}{6} (-f(-2 h)+4
f(-h)+4 f(h)-f(2 h))+\frac{\ep 
  (f(-2 h)-8 f(-h)+8 f(h)-f(2 h))}{12 h}
\nonumber\\
&&{}
+\frac{\ep^2 (f(-2 h)-f(-h)-f(h)+f(2 h))}{6 h^2}
+\frac{\ep^3 (-f(-2 h)+2 f(-h)-2 f(h)+f(2 h))}{12
  h^3}
\nonumber\\
&&{}
+O\left(\epsilon ^4\right)\,.
\label{eq:interEx}
\end{eqnarray}
Now, taking into account that 
\begin{equation}
f(\ep) = f(0)+f'(0)\epsilon+\frac{1}{2}  f''(0)\epsilon ^2+\frac{1}{6}
f^{(3)}(0) \epsilon ^3+O\left(\epsilon ^4\right)\,,
\label{eq:serEx}
\end{equation}
expanding in $h$ the coefficients in front of powers of $\ep$ in Eq.~\eqref{eq:interEx},
\begin{eqnarray}
\frac{1}{6} (-f(-2 h)+4
f(-h)+4 f(h)-f(2 h)) & =& f(0)-\frac{1}{6} f^{(4)}(0) h^4+O\left(h^5\right)\,,
\nonumber\\
\frac{ 1}{12 h}
(f(-2 h)-8 f(-h)+8 f(h)-f(2 h))  & =& f'(0)-\frac{1}{30} f^{(5)}(0) h^4+O\left(h^5\right)\,,
\nonumber\\
\frac{1}{6 h^2}(f(-2 h)-f(-h)-f(h)+f(2 h)) & =& \frac{f''(0)}{2}+\frac{5}{24} f^{(4)}(0) h^2+O\left(h^4\right)\,,
\nonumber\\
\frac{1}{12
	h^3} (-f(-2 h)+2 f(-h)-2 f(h)+f(2 h)) & =& \frac{1}{6} f^{(3)}(0)+\frac{1}{24} f^{(5)}(0) h^2+O\left(h^4\right)\,,
\end{eqnarray}
and comparing them with the expansion coefficients in Eq.~\eqref{eq:serEx}, we get an estimate for the accuracy of our interpolation.
If the function $f(\ep)$ has a pole of order $m$ in $\ep$, we can simply interpolate the function $g(\ep) = \ep^m f(\ep)$ instead.

Returning to our example in Eq.~\eqref{eq:BruteEx}, we see that, while previously we had to solve the system of Eqs.~\eqref{eq:BruteEx1}--\eqref{eq:BruteEx4}, now we only need to solve Eq.~\eqref{eq:BruteEx} four times, for $\ep = -2h, -h, h, 2h$. Despite the small value of $h$, this approach will be computationally preferable for sufficiently large $n$ values, as the integer numbers in the coefficients $c_n^{(3)}$ grow as $n^8$.  More importantly, this approach allows for efficient parallelization. The calculations at each lattice node are independent, whereas, in the original method, the expansion coefficients depend on each other making their evaluation non-parallelizable.

The error in our approach can be estimated quite accurately. It comes from two sources. The first one is the interpolation error, and the second one is the error associated with the truncation of the Frobenius power series at the evaluation point. In general, we estimate our error as
\begin{equation}
\text{max}\left(h^{2n - 2\lfloor  k/2 \rfloor} , \Delta_{\text{Frob}}\right)\,,
\end{equation}
where $h$ is the lattice step, $2n$ is the number of lattice elements, $k$ is the number of terms in the $\ep$ expansion, and $\Delta_{\text{Frob}}$ is the error arising from the truncation of the Frobenius generalized power series.

\subsection{Algorithm summary and example}
\label{sec:AlgorithmSummary}

Having presented a detailed description of the various steps of our calculation procedure, let us now give its summary. The algorithm takes as input the target function, the evaluation point, the required precision and the required number of terms in the $\ep$ expansion. Additionally, it requires the differential-equation system in Pfaffian form which the target function satisfies. However, these differential-equation systems are already built into our software package PrecisionLauriecella,\footnote{See Appendix \ref{appendix::PrecisionLauricella}, for its usage.} so that the users do not need to worry about them unless they wish to add their own functions. The output of the algorithm is the numerical expansion of the target function at the evaluation point up to the required order in the parameter $\ep$. The algorithm then proceeds in the following steps:
\begin{enumerate}   
	\item Starting from the system of differential equations in Pfaffian form and the given evaluation point, derive a differential equation system in one variable.
	
	\item Fix the boundary conditions based on a known basis of functions for the differential equation system together with their expansions at the origin.  Also, determine the order of the pole in $\ep$ if one exists.
	
	\item Determine the path from the origin to the evaluation point by constructing an analytic continuation.
	
	\item Based on the required expansion order in $\ep$, choose a set of  $\ep$ lattice  points  at which the evaluations will be performed. The lattice step is determined on the basis of the required accuracy.
	
	\item Derive boundary conditions and recurrence relations for the Frobenius series coefficients  for all regions along  the analytic-continuation path and all $\ep$ values chosen in the previous step.
	
	\item Using the derived recurrence relations, numerically evaluate the $U$ matrix at the origin, the evaluation point and each matching point along the analytic-continuation path. Do that for each $\ep$ value.
	
	\item Finally, calculate the value of the target function at the evaluation point for each $\ep$ value and perform the reconstruction of its Laurent series expansion in $\ep$ using Lagrange interpolation polynomials.
\end{enumerate}

\begin{figure}[ht]
	\centering
	\includegraphics[width=0.5\textwidth]{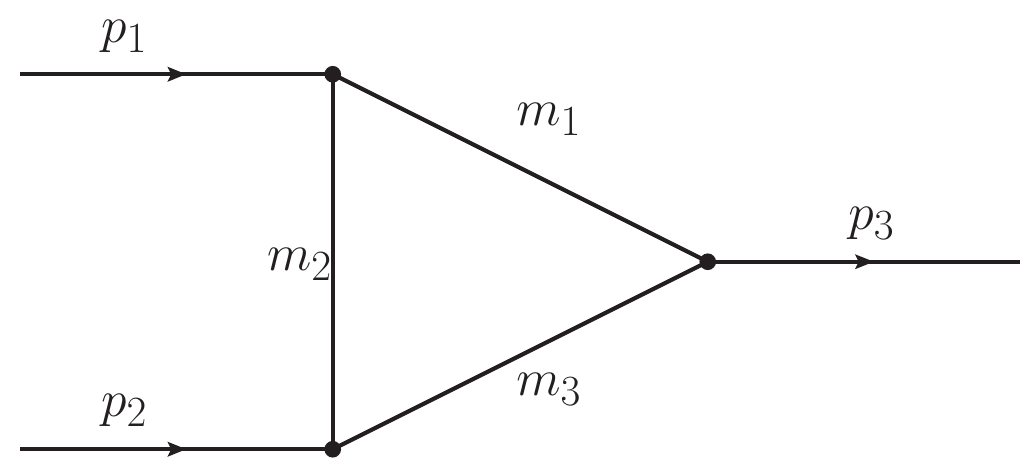}
	\caption{One-loop triangle with six independent scales.}
	\label{fig:triangle}
\end{figure}

As an example of the described procedure, let us consider the evaluation of the following function:
\begin{equation}
F_1\left(\frac{1}{2};1,\ep;\frac{3}{2};\frac{4}{3},\frac{7}{4}\right)\,,
\end{equation}
which appears in the calculation of Feynman diagrams \cite{Tarasov:2022clb} and is sufficiently simple to serve as good example. More specifically, this function arises in the computation of the one-loop triangle with six independent scales, shown in Fig.~\ref{fig:triangle}.

Using the Diogenes \cite{Bezuglov:2023owj} software package, we can obtain the expansion of this function in terms of multiple polylogarithms,
\begin{eqnarray}
F_1\left(\frac{1}{2};1,\ep;\frac{3}{2};x,y\right) &=& \frac{G\left(-1,\sqrt{x}\right) - G\left(1,\sqrt{x}\right)}{2 \sqrt{x}} + \frac{\varepsilon}{2\sqrt{x}} \Bigg( -G\left(-1,\sqrt{x}\right) \left(G\left(-1,\sqrt{y}\right) + G\left(1,\sqrt{y}\right)\right) \nonumber\\
&&{}+ G\left(1,\sqrt{y}\right) \left(G\left(-\sqrt{y},\sqrt{x}\right) - G\left(\sqrt{y},\sqrt{x}\right)\right) + \dots \Bigg) + \mathcal{O}(\ep^2)\,,
\end{eqnarray}
and evaluate the latter numerically at the required point using tools like GiNaC \cite{Vollinga:2004sn} or HandyG \cite{Naterop:2019xaf}. This will provide us with an independent check of our calculation procedure.

Choosing the basis of functions as
\begin{equation}
J = \left\{F_1, \, x\frac{\partial}{\partial x}F_1, \, y\frac{\partial}{\partial y}F_1 \right\}\,,
\end{equation}
the corresponding Pfaffian differential-equation system takes the form
\begin{equation}
dJ = \left( M_x \, dx + M_y \, dy \right) J\,,
\end{equation}
where
\begin{eqnarray}
M_x &=& \left(
\begin{array}{ccc}
0 & \frac{1}{x} & 0 \\
\frac{1}{2 - 2x} & \frac{-3x^2 + (3 - 2\varepsilon)xy + x + (2\varepsilon - 1)y}{2(x - 1)x(x - y)} & \frac{y - 1}{(x - 1)(x - y)} \\
0 & \frac{\varepsilon y}{x^2 - xy} & \frac{1}{y - x} \\
\end{array}
\right)
\,,
\nonumber\\
M_y &=& \left(
\begin{array}{ccc}
0 & 0 & \frac{1}{y} \\
0 & \frac{\varepsilon}{x - y} & \frac{x}{y(y - x)} \\
\frac{\varepsilon}{2 - 2y} & -\frac{\varepsilon (x - 1)}{(y - 1)(x - y)} & \frac{x + y}{2xy - 2y^2} - \frac{\varepsilon}{y - 1} \\
\end{array}
\right)\,.
\end{eqnarray}
Next, selecting the path $x \rightarrow \frac{t}{3}$ and $y \rightarrow \frac{7t}{16}$, the $M_t$ differential system in the variable $t$ is defined by the matrix
\begin{equation}
M_t = \left(
\begin{array}{ccc}
0 & \frac{1}{t} & \frac{1}{t} \\
\frac{1}{6 - 2t} & -\frac{3(t - 1)}{2(t - 3)t} & \frac{1}{3 - t} \\
\frac{7\varepsilon}{32 - 14t} & \frac{7\varepsilon}{16 - 7t} & \frac{16 - 7(2\varepsilon + 1)t}{2t(7t - 16)} \\
\end{array}
\right)\,,
\label{eq:MTEx}
\end{equation}
and the boundary conditions vector by $b = \{1, 0, 0\}$.
\begin{figure}[ht]
	\centering
	\includegraphics[width=0.5\textwidth]{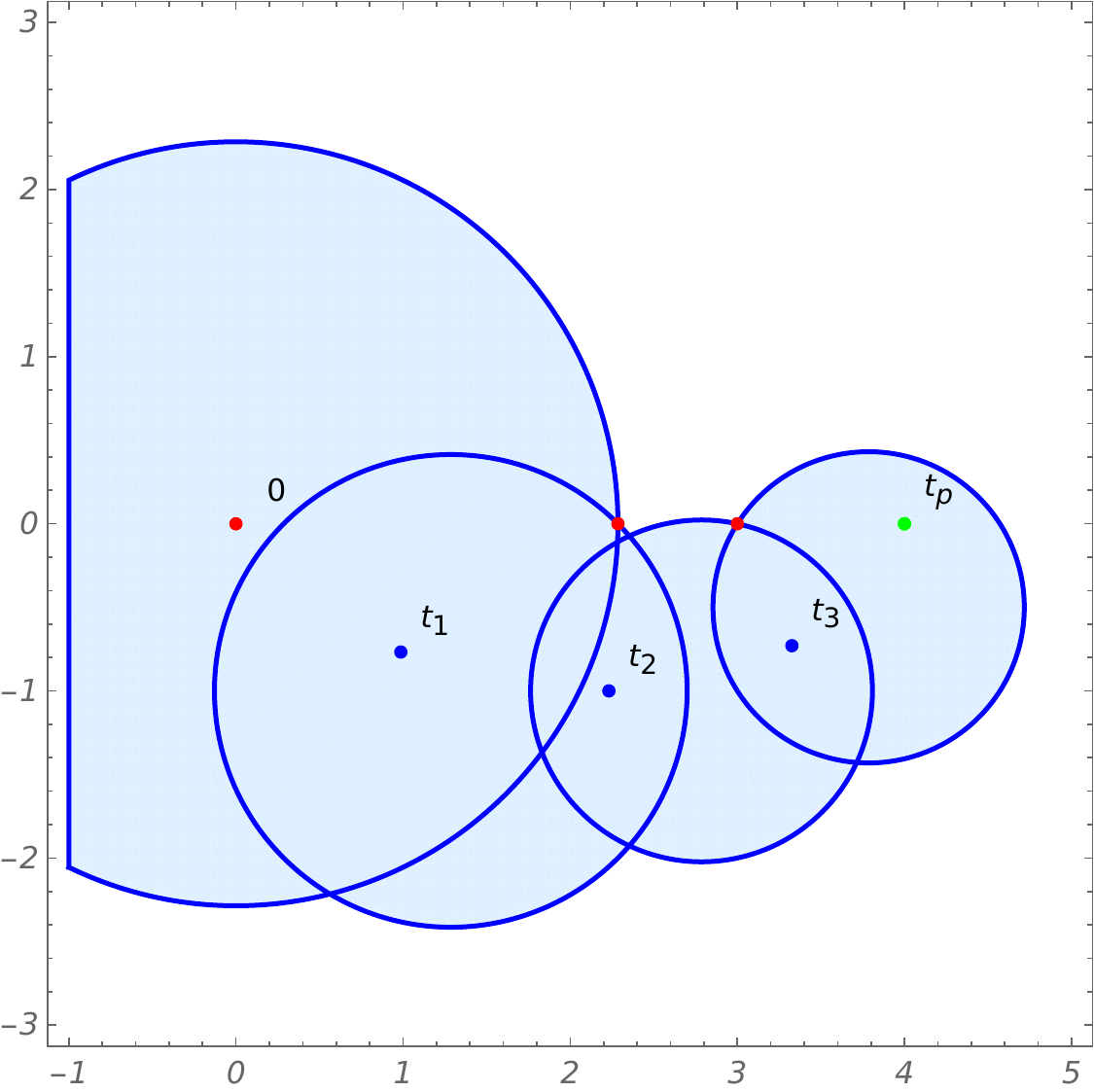}
	\caption{This figure illustrates the path of the analytic continuation of the system in Eq.~\eqref{eq:MTEx} from the origin $0$ to the evaluation point $t_p = 4$. Red points denote singularities, shaded circles represent convergence regions around the points $\left\{0, \frac{9}{7} - i, \frac{39}{14} - i, \frac{53}{14} - \frac{i}{2}\right\}$, blue points indicate the gluing points of these regions, and the green point is our evaluation point.}
	\label{fig:path}
\end{figure}
Thus, the problem reduces to solving this differential-equation system at the point $t_p = 4$. For that, we first need to construct an analytic-continuation path.  As was already discussed in Section~\ref{sec:AnalyticalContinuation}, this path is not unique.  Our software package selects the path shown in Fig.~\ref{fig:path}. Consequently, the analytic continuation proceeds through the regions centered at the points $\left\{0, \frac{9}{7} - i, \frac{39}{14} - i, \frac{53}{14} - \frac{i}{2}\right\}$, and the value of the considered function at the evaluation point $t_p$ is assembled in the following way:
\begin{eqnarray}
F_1\left(\frac{1}{2};1,\ep;\frac{3}{2};\frac{4}{3},\frac{7}{4}\right)&=&
U^{\frac{53}{14} - \frac{i}{2}}(t_p) \left(U^{\frac{53}{14} - \frac{i}{2}}(t_3)\right)^{-1}
U^{\frac{39}{14} - i}(t_3) \left(U^{\frac{39}{14} - i}(t_2)\right)^{-1}
\nonumber\\
&&{}\times
U^{\frac{9}{7} - i}(t_2) \left(U^{\frac{9}{7} - i}(t_1)\right)^{-1}
U^{0}(t_1) \left(U^{0}(0)\right)^{-1} b\, ,
\end{eqnarray}
where $U^{S}(x)$ denotes the fundamental series solutions matrix obtained as a series expansion around the point $S$ and evaluated at the point $x$, see Section~\ref{sec:SeriesSolution}. The final step is the reconstruction of the $\ep$ expansion. For that, we use the procedure described in Section~\ref{sec:EpExpansion}. For example, if we want the expansion up to $\ep^3$ with a precision of 30 decimal places, we can choose a set of four lattice points with a step size of $h = 10^{-18}$: $\ep \in \left\{-\frac{2}{10^{18}}, -\frac{1}{10^{18}}, \frac{1}{10^{18}}, \frac{2}{10^{18}}\right\}$. Notice that we specifically choose such small step size to achieve the required precision with some extra "buffer" or "stock," which generally allows for more reliable results. Combining everything together gives us the following result:
\begin{eqnarray}
  \lefteqn{F_1\left(\frac{1}{2};1,\ep;\frac{3}{2};\frac{4}{3},\frac{7}{4}\right)}
\nonumber\\  
&=&(1.14051899445141952129664138232 - 1.36034952317566338794555869323\,i)
\nonumber\\&&{}
- (1.93816954384142983458363185442 + 1.50595641724256995525115087323\,i) \, \varepsilon
\nonumber\\&&{}
- (1.67642008095711823380650561964 - 2.07761091570717412690937916205\,i) \, \varepsilon^2
\nonumber\\&&{}
+ (1.64228238234018020089070332528 + 1.43969305215049203442016005240\,i) \, \varepsilon^3\,.\qquad
\end{eqnarray}

\section{Conclusion}
\label{sec:Conclusion}

In this work, we presented a method for high-precision numerical evaluations of Lauricella functions whose indices are linearly dependent on some parameter $\ep$, in terms of their Laurent series expansions about $\ep = 0$. The latter are based on finding analytic continuations of these functions in terms of Frobenius generalized power series. Being one-dimensional, these series are much more suited for high-precision numerical evaluations than multi-dimensional series arising in approaches to analytic continuation based on re-expansions of hypergeometric series or Mellin--Barnes integral representations. Another advantage is provided by our treatment of the $\ep$ dependence in Laurent expansions of Lauricella functions, offering another significant acceleration in terms of computational efficiency. Our treatment of the $\ep$ dependence is also well suited for parallel computations, making it ideal for large-scale calculations that require high performance and scalability.

\begin{figure}[ht]
	\centering
	\includegraphics[width=0.9\textwidth]{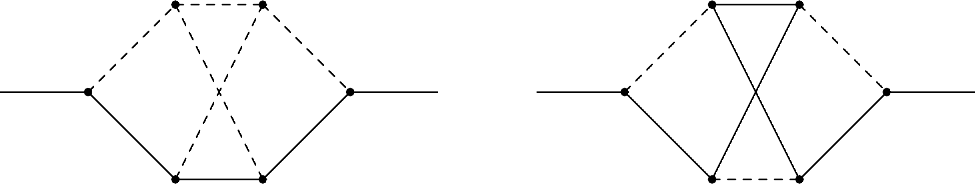}
	\caption{Examples of Feynman integrals on which we tested our method. Dashed lines denote massless propagators, and thick lines represent massive propagators.}
	\label{fig:exPheynman}
\end{figure}

The advocated calculation procedure was implemented in the \texttt{PrecisionLauricella} package, written in Wolfram Mathematica language. The latter provides a universal and efficient tool for numerical high-precision evaluations of Lauricella and similar hypergeometric functions. To test our package, we used, whenever possible, alternative packages for evaluations of hypergeometric functions, such as those presented in Refs.~\cite{Ananthanarayan:2021bqz,Bera:2024hlq} and Refs.~\cite{Bera:2023pyz,Bezuglov:2023owj}. In the latter case, the provided expansions of hypergeometric functions in terms of multiple polylogarithms can be evaluated with high precision using other known mathematical packages such as those presented in Refs.~\cite{Vollinga:2004sn, Naterop:2019xaf}.

There are several possible directions for future research. First, the presented approach can be extended to include additional classes of hypergeometric functions of several variables, such as the Kampé de Fériet, Horn and GKZ $\mathcal{A}$-hypergeometric functions. In all these cases, an automation for the derivation of the corresponding Pfaffian differential equation systems is required. Second, our approach allows us to build one-dimensional generalized power series representations of the considered functions in different regions of $\mathbb{C}^n$, where $n$ is the number of variables in the corresponding hypergeometric functions, and to compare them with multi-dimensional series representations coming from re-expansions of hypergeometric series, Mellin--Barnes integral representations or with GKZ $\Gamma$ series. These regions can be chosen to cover the whole $\mathbb{C}^n$ space. Moreover, one can study the dependence of the convergence of the provided series solutions on different $\mathbb{C}^n$ coverings and their refinements. The knowledge of such generalized power series representations for  hypergeometric functions will also allow us to perform computations of the corresponding monodromy groups and to perform comparisons with already available results on the monodromies of hypergeometric functions \cite{beukers2013monodromy,goto2017monodromy,matsumoto2017monodromy,goto2022monodromy}.   Finally, we plan to apply the presented approach directly to the computation of Feynman integrals. Preliminary considerations of the Feynman integrals shown in Fig.~\ref{fig:exPheynman} have yielded promising results.  The calculation of a system of 53 master integrals with six terms in the $\ep$ expansions took only several minutes. However, Feynman integrals pose additional challenges related to their more complex boundary conditions and the derivations of the corresponding Pfaffian differential systems.  Still, the results for master integrals in terms of one-dimensional generalized power series representations valid in any region of kinematic parameter space are very desirable. This is especially the case when we take into account their subsequent numerical integration over the phase space. These representations allows us, for example, to efficiently train neural networks describing them and to use the latter in adaptive Monte-Carlo phase space integrations with normalizing flows \cite{NormFlows1,NormFlows2,NormFlows3}.

These developments will form the core of our future research, and we expect that the continued refinement of this approach will open up new possibilities in both theoretical physics and applied mathematics, particularly in the computation of Feynman integrals and hypergeometric functions.

\section*{Acknowledgments}
We would like to thank V.V.Bytev, R.N.Lee and A.V.Kotikov  for interesting and stimulating discussions. The work of A.O. was supported by the Russian Science Foundation through Grant No.~20-12-00205.
The work of M.B. and B.K. was supported by the German Research Foundation DFG through Grant No.~KN 365/16-1.
The work of O.V. was supported by the DFG Research Unit FOR 2926 through Grant No.\ KN 365/13-2.

\appendix

\section{PrecisionLauricella package}
\label{appendix::PrecisionLauricella}
The \texttt{PrecisionLauricella} package can be freely downloaded from the bitbucket repository \url{https://bitbucket.org/BezuglovMaxim/precisionlauricella-package/src/main/}. A comprehensive documentation of it may be found in Ref.~\cite{cpc}. The entire package consists of the single file PrecisionLauricella.wl and, provided the path is set correctly, is loaded with the command 
\begin{lstlisting}[language=Mathematica]
<< PrecisionLauricella`
\end{lstlisting}
\begin{table}[h]
	\centering
	\begin{tabular}{|c|c|c|c|c|c|c|}
		\hline
		precision & $\ep^0$ & $\ep^2$ & $\ep^4$ & $\ep^6$ & $\ep^8$ & $\ep^{10}$ \\ \hline
		20    & $0.41$ & $0.61$ & $1.11$ & $1.35$ & $3.18$ & $10.27$    \\ \hline
		60    & $1.24$ & $1.64$ & $3.13$ & $3.61$ & $6.4$ & $14.05$    \\ \hline
		100    & $2.8$ & $3.57$ & $7.57$ & $9.45$ & $14.71$ & $24.2$   \\ \hline
		200    & $11.83$ & $17.03$ & $39.59$ & $47.77$ & $68.48$ & $85.8$   \\ \hline
	\end{tabular}
	\caption{Average time in seconds required to expand the $F_1\left(\frac{1}{2};1,\ep;\frac{3}{2};\frac{4}{3},\frac{7}{4}\right)$ function using 16 parallel kernels. The horizontal axis shows the number of terms in $\ep$, and the vertical axis shows the calculation accuracy in digits.}
	\label{tab:example1}
\end{table}
\begin{table}[h]
	\centering
	\begin{tabular}{|c|c|c|c|c|c|c|}
		\hline
		precision & $\ep^0$ & $\ep^2$ & $\ep^4$ & $\ep^6$ & $\ep^8$ & $\ep^{10}$ \\ \hline
		20    & $0.51$ & $0.73$ & $1.16$ & $1.43$ & $3.32$ & $9.93$    \\ \hline
		60    & $1.23$ & $2.27$ & $3.97$ & $5.24$ & $8.46$ & $16.27$    \\ \hline
		100    & $2.75$ & $5.15$ & $10.52$ & $12.37$ & $19.17$ & $33.42$   \\ \hline
		200    & $11.8$ & $30.09$ & $55.25$ & $69.18$ & $98.68$ & $127.95$   \\ \hline
	\end{tabular}
	\caption{Average time in seconds required to expand the $F_1\left(\frac{1}{2};1,\ep;\frac{3}{2};\frac{4}{3},\frac{7}{4}\right)$ function using 8 parallel kernels. The horizontal axis shows the number of terms in $\ep$, and the vertical axis shows the calculation accuracy in digits.}
	\label{tab:example2}
\end{table}
The universal function for the numerical expansions of Lauricella functions in Laurent series in the parameter $\ep$ is \texttt{NExpandHypergeometry}. This function takes three arguments. The first one is a hypergeometric function that is to be expanded. The second one is the list $\{\ep, k\}$, where the first argument $\ep$ is the variable with respect to which the expansion is to be performed, and the second one is the required number of terms in the $\ep$ expansion. The last argument specifies the accuracy of the result with the desired number of decimal places. For example,
\begin{mmaCell}[moredefined={NExpandHypergeometry}]{Input}
NExpandHypergeometry[AppellF1[\mmaFrac{1}{2},1,\(\varepsilon\),\mmaFrac{3}{2},\mmaFrac{4}{3},\mmaFrac{7}{4}],\{\(\varepsilon\),3\},30]
\end{mmaCell}
\begin{mmaCell}{Output}
(1.14051899445141952129664138232 - 1.36034952317566338794555869323 I)
- (1.93816954384142983458363185442 + 1.50595641724256995525115087323 I) \(\varepsilon\)
- (1.67642008095711823380650561964 - 2.07761091570717412690937916205 I) \mmaSup
+ (1.64228238234018020089070332528 + 1.43969305215049203442016005240 I) \mmaSup

\end{mmaCell}

For those functions that are not included by default in Wolfram Mathematica, we introduce our own notations:  \texttt{AppellF2}, \texttt{AppellF3}, \texttt{LauricellaFA}, \texttt{LauricellaFB} and \texttt{LauricellaFD}.  The arguments of these functions are the same as in their definitions. In the case of Lauricella functions, numbered indices and variables are collected into lists.  More details about the package functionality can be found in the Mathematica notebook with examples.
\begin{table}[h]
	\centering
	\begin{tabular}{|c|c|c|c|c|c|c|}
		\hline
		precision & $\ep^0$ & $\ep^2$ & $\ep^4$ & $\ep^6$ & $\ep^8$ & $\ep^{10}$ \\ \hline
		20    & $1.93$ & $2.38$ & $3.54$ & $4.64$ & $6.93$ & $14.24$    \\ \hline
		60    & $4.42$ & $4.84$ & $8.33$ & $11.34$ & $14.14$ & $23.14$    \\ \hline
		100    & $7.3$ & $8.64$ & $12.65$ & $15.13$ & $26.77$ & $38.03$   \\ \hline
		200    & $16.42$ & $18.38$ & $30.2$ & $38.15$ & $51.26$ & $63.95$   \\ \hline
	\end{tabular}
	\caption{Average time in seconds required to expand the $F_2\left(1,\frac{2}{3}\ep,1,1+\frac{3}{2}\ep,1-\frac{15}{7}\ep;\frac{3}{2},4\right)$ function using 8 parallel kernels. The horizontal axis shows the number of terms in $\ep$, and the vertical axis shows the calculation accuracy in digits.}
	\label{tab:example3}
\end{table}
\begin{table}[h]
	\centering
	\begin{tabular}{|c|c|c|c|c|c|c|}
		\hline
		precision & $\ep^0$ & $\ep^2$ & $\ep^4$ & $\ep^6$ & $\ep^8$ & $\ep^{10}$ \\ \hline
		20    & $1.92$ & $3.57$ & $5.59$ & $6.93$ & $10.75$ & $18.9$    \\ \hline
		60    & $3.68$ & $7.29$ & $13.98$ & $17.23$ & $25.4$ & $35.31$    \\ \hline
		100    & $6.37$ & $12.5$ & $21.2$ & $26.96$ & $42.43$ & $55.53$   \\ \hline
		200    & $17.16$ & $33.38$ & $51.8$ & $69.88$ & $87.55$ & $107.72$   \\ \hline
	\end{tabular}
	\caption{Average time in seconds required to expand the $F_D^{(3)}(\frac{1}{2}-\ep; 1,\ep,\ep; 1+2\ep; \frac{4}{3},\frac{3}{4},\frac{8}{5})$ function using 8 parallel kernels. The horizontal axis shows the number of terms in $\ep$, and the vertical axis shows the calculation accuracy in digits.}
	\label{tab:example4}
\end{table}
\begin{table}[h]
	\centering
	\begin{tabular}{|c|c|c|c|c|c|c|}
		\hline
		precision & $\ep^0$ & $\ep^2$ & $\ep^4$ & $\ep^6$ & $\ep^8$ & $\ep^{10}$ \\ \hline
		20    & $11.47$ & $15.53$ & $24.96$ & $29.22$ & $41.59$ & $51.30$    \\ \hline
		60    & $16.03$ & $26.20$ & $45.88$ & $55.91$ & $85.11$ & $92.26$    \\ \hline
		100    & $24.56$ & $38.83$ & $74.62$ & $92.84$ & $121.96$ & $143.40$   \\ \hline
		200    & $49.43$ & $97.83$ & $187.28$ & $215.97$ & $282.98$ & $334.059$   \\ \hline
	\end{tabular}
	\caption{Average time in seconds required to expand the $F_B^{(3)}(1+4\ep,3\ep,\ep; 1+2\ep,1,\ep; 1+\ep; \frac{4}{3},-\frac{3}{4},\frac{8}{5})$ function using 8 parallel kernels. The horizontal axis shows the number of terms in $\ep$, and the vertical axis shows the calculation accuracy in digits.}
	\label{tab:example5}
\end{table}

The computation time for certain functions is of particular interest. Examples of such test calculations are provided in Tables \ref{tab:example1}--\ref{tab:example5}. The test calculations were performed on a standard laptop with a 12th Gen Intel® Core™ i7-12700H × 20 processor. From these tables, several conclusions about the program's performance can be drawn. First, as may be seen from Tables \ref{tab:example1} and \ref{tab:example2}, a higher number of parallel Mathematica kernels provides an advantage in the cases when higher numerical precision and larger numbers of terms in the $\ep$ expansion are required. This is easily understood by observing that, for both complex and simple tasks, the program finds the same analytic continuation and recurrence relations for the expansion coefficients in the Frobenius series. This takes time and cannot be efficiently parallelized. However, in complex tasks, more time is spent for solving recurrence relations for the expansion coefficients, and this part can be effectively parallelized, as described in Section~\ref{sec:EpExpansion}. Second, as the number of terms in the $\ep$ expansion grows, the program’s execution time also grows linearly with their number. This is a direct consequence of our approach for treating $\ep$ dependencies, see  Section~\ref{sec:EpExpansion}. Finally, comparing Tables \ref{tab:example1}, \ref{tab:example2} with \ref{tab:example4}, \ref{tab:example5}, it is seen that the time required for computations of hypergeometric functions using the Frobenius method depends not as much as on the number of their variables as on the complexity of the functions themselves. In our approach, the latter is related to the number of linearly independent solutions of the corresponding systems of differential equations, that is the functional basis size. For example, the functional basis size is three for $F_1$ functions, four for $F_2$ and $F_D^{(3)}$ functions, and eight for $F_B^{(3)}$ functions. 

\bibliographystyle{hieeetr}
\bibliography{litr}

\end{document}